# Water, air and fire at work in Hero's machines

**Amelia Carolina Sparavigna**
Dipartimento di Fisica, Politecnico di Torino
Corso Duca degli Abruzzi 24, Torino, Italy

*Known as the Michanikos, Hero of Alexandria is considered the inventor of the world's first steam engine and of many other sophisticated devices. Here we discuss three of them as described in his book "Pneumatica". These machines, working with water, air and fire, are clear examples of the deep knowledge of fluid dynamics reached by the Hellenistic scientists.*

Hero of Alexandria, known as the Mechanicos, lived during the first century in the Roman Egypt [1]. He was probably a Greek mathematician and engineer who resided in the city of Alexandria. We know his work from some of writings and designs that have been arrived nowadays in their Greek original or in Arabic translations. From his own writings, it is possible to gather that he knew the works of Archimedes and of Philo the Byzantian, who was a contemporary of Ctesibius [2].

It is almost certain that Heron taught at the Museum, a college for combined philosophy and literary studies and a religious place of cult of Muses, that included the famous Library. For this reason, Hero claimed himself a pupil of Ctesibius, who was probably the first head of the Museum of Alexandria. Most of Hero's writings appear as lecture notes for courses in mathematics, mechanics, physics and pneumatics [2]. In optics, Hero formulated the Principle of the Shortest Path of Light, principle telling that if a ray of light propagates from a point to another one within the same medium, the followed path is the shortest possible. After a thousand of years, Ibn al-Haytham expanded the principle to reflection and refraction [3]: it was only in 1662, that Fermat stated the principle in its modern form. In mathematics, Hero described an iterative method for the computing square roots. Today, his name is associated with the formula for finding the area of a triangle from the lengths of its sides.

Hero, the Michanikos, is celebrated for his activity as an engineer: in his books, he describes several machines and automata working under the action of water and steam by means of an excellent use of siphons and valves. It is noticeable that many of the devices he is describing are incorporating some sorts of feedback control systems. The University of Rochester publishes one of his books, "Pneumatica", in its web library on Steam Engines. The book contains the detailed descriptions of machines, descriptions translated from the Greek original, illustrated by drawings that, according to Bennet Woodcroft, Editor of this English translation of 1851, have been made expressly for the English edition from the best examples from previous editions.

"Pneumatica" contains a description of a steam-powered device, called "aeolipile", or also the "Hero Engine", which is considered the first recorded steam engine or reaction steam turbine. The name comes from the words "Aeolus" and "pila", that is "the ball of Aeolus", Aeolus being the god of winds. Vitruvius mentioned in his "De Architectura" that aeolipile was created earlier than the Hero's description and probably it was a device invented by Ctesibius (285–222 BC). In any case, it means that steam devices were developed almost two millennia before the industrial revolution.

Aeolipile is a hollow sphere that can rotate about an axis passing through antipodal points, because of steam flowing out through two bend pipes placed at its equator. Figure 1 shows the device (the figure was prepared using as model that of Bennet Woodcroft's Edition). Let us follow the Hero's description, in the English edition, at the section entitled "Place a cauldron over a fire: a ball shall revolving on a pivot". A fire is lighted under a cauldron containing water and covered by a lid; a bent tube is passing through the lid, the extremity of the tube being fitted into a hollow ball. Opposite to the extremity of this tube, a pivot is placed, resting on the lid. The ball has two bent pipes, communicating with the ball and placed at the opposite extremities of a diameter, and bent in opposite directions. As the cauldron gets hot, the steam, entering the ball through the tube, passes out through the bent pipes and causes the ball to revolve.

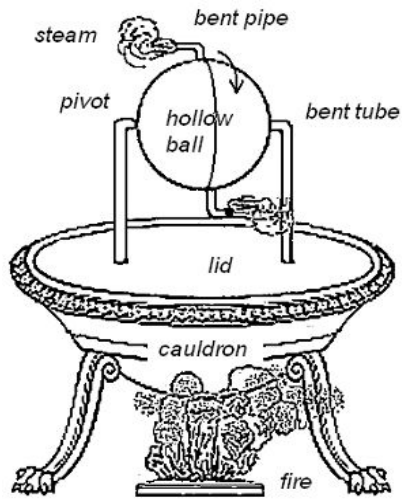

Figure 1 Aeolipile or the Hero's Machine. Heating the water in the cauldron with a fire, the steam enters the hollow ball through the bent tube and passes out through the two bent pipes, causing the ball to revolve.

There is another device proposed by Hero, which is very simple, nevertheless quite interesting. This is a device based on the Bernoulli Effect. It is described in the section entitled "A jet of steam supporting a sphere". Heron is telling that small balls can be supported aloft in the following manner. Underneath a cauldron containing water and covered at the top, a fire is lighted (Figure 2). From the covering lid, a tube runs upwards, at the extremity of which and communicating with it, is a hollow hemisphere. If we put a light ball into the hemisphere, it will be found that the steam from the cauldron, rising through the tube, lifts the ball so that it is suspended.

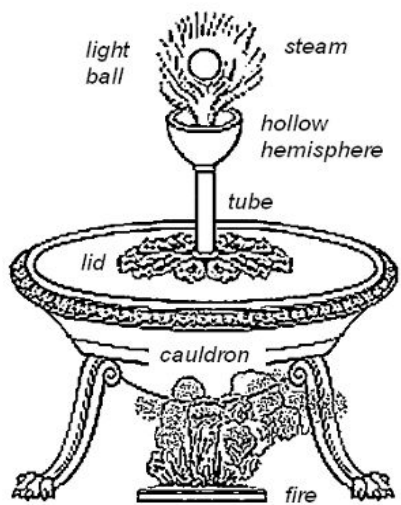

Figure 2. The ball is floating suspended by the Bernoulli Effect in the flowing steam

We can try to reproduce this effect with a ping-pong ball and suspend it in the air flow from a hair dryer. When the air streams symmetrically around the ball in the upward direction, air drag and gravity point in opposite directions, and drag balances gravity. A small perturbation of the ball position will produce the Magnus Effect, probably observed by Hero, but not described in his text.

"Pneumatica" contains quite complex devices too. One is that described in the section "Temple doors opened by fire on an altar". This section shows how to construct a mechanism able to open and close automatically the doors of a small temple, in front of the amazed worshippers. On lighting a fire on a near altar, the doors shall open, and shut again when the fire is extinguished.

The temple and altar stand on a pedestal concealing the mechanism (see Figure 3). The adapted description is the following. Under the altar, there is a globe. A tube starts from a cavity under the lid of the altar, enters the globe ending at nearly to its centre. The tube must be soldered into the globe, in which a bent siphon is placed too. The hinges of the doors extend downwards and turn freely on pivots in the floor. From the hinges let two chains be attached, by means of a pulley, to a hollow suspended vessel. Other two chains, wound upon the hinges in the opposite direction to the former: these chains are attached by means of a pulley, to a leaden weight. Descending this weight, the doors will be shut. Let us now examine the siphon: the outer leg of the siphon ends into the suspended vessel. Through a hole, P, which must be carefully closed afterwards, pour water into the globe enough to fill one half of it. When the fire on the altar is very hot, the pressure of air in the altar increases. The pressure in the globe increases too, driving out some liquid contained in it into the vessel, through the siphon. The vessel, descending under the weight, will tighten the chains and open the doors. When the fire is extinguished, the pressure decreases, the water flows back into the globe through the bent siphon from the suspended vessel. The vessel lightened, the weight suspended will preponderate and shut the doors.

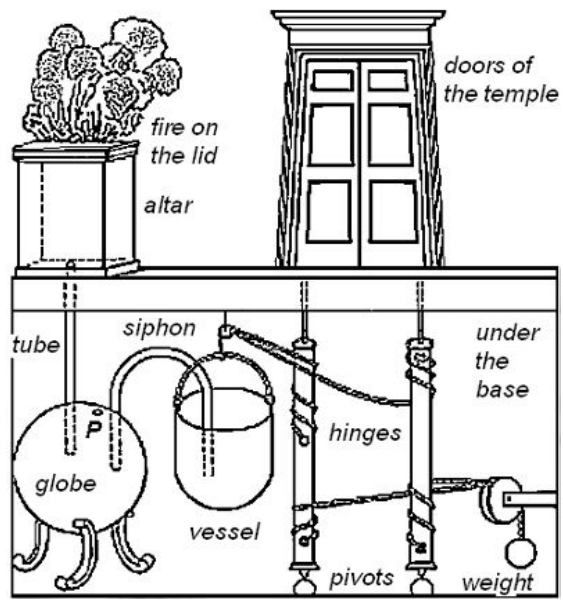

Figure 3. When the fire on the altar has grown hot, the pressure in the globe increases, driving out some liquid contained there through the siphon into a suspended vessel. The vessel, descending under the weight, will tighten the chains and open the doors. When the fire is extinguished, the pressure decreases, the water flows back into the globe through the bent siphon from the suspended vessel. The weight suspended will preponderate and shut the doors.

At the end of the description, Hero suggests to use quicksilver, for a better result. This comment seems to tell us that he prepared some models with water and quicksilver and compared the resultant behaviors with the two working fluids.

Many other mechanisms and automations, working with water, air and fire, are described in "Pneumatica", each of them very interesting and intriguing. Mainly proposed and produced for worship places and theatres, many of these objects had no other function than that of people entertainment. Hero did not invent all these machines, but probably reported his own work with that of past scientists. In any case, the knowledge of fluid dynamics underlying their description is amazing.

We could ask ourselves, how old and deep was this knowledge? It is difficult to answer due to an obvious lack of documents. According to some scholars, Hero represents the top of Hellenistic science, a science which reached heights not achieved by the Classical Age of Science. Let us conclude with some remarks by Lucio Russo [6], who wrote the book "*The Forgotten Revolution: how science was born in 300 BC and why it had to be reborn*". He tells that the Hellenistic science went even further than we can ordinarily thought, but that the results of this science were lost with the Roman conquest. In the 16th Century, as the ancient texts started once again to be available in Europe, the legacy of Hellenistic science served as the base of the scientific revolution. Machines that Hero used to amaze people in the temples started a new life in the subsequent industrial revolution.